\documentclass[12pt]{article}
\pdfoutput=1
\usepackage{amsmath,amssymb,amsfonts,amsthm,bm,bbm,cancel,wasysym}
\usepackage{epsfig,graphics,graphicx,epstopdf,caption,subcaption}
\usepackage[numbers,sort&compress]{natbib}
\graphicspath{{Charts/}}
\usepackage{array,booktabs,colortbl,colordvi,multirow}
\usepackage{colordvi,color,xcolor}
\usepackage{hyperref}
\usepackage{rotating}
\usepackage{comment}
\usepackage{feynmf}

\usepackage[symbol]{footmisc}
\renewcommand{\thefootnote}{\fnsymbol{footnote}}

\begin{document}

\begin{center}

{\Large {\bf LHC-friendly freeze-in dark matter via Higgs portal}}\\

\vspace*{0.75cm}

{Xinyue Yin,$^{a}$ Shuai Xu,$^{b}$ and Sibo Zheng$^{a,}$\footnote{Contact author: sibozheng.zju@gmail.com}}

\vspace{0.5cm}
{$^{a}$Department of Physics, Chongqing University, Chongqing 401331, China \\
$^{b}$School of Physics and Telecommunications Engineering, Zhoukou Normal University, Henan 466001, China }
\end{center}
\vspace{.5cm}

\begin{abstract}
\noindent
It is known that single-field freeze-in dark matter barely leaves footprints in dark matter direct detection and collider experiments. 
This situation can be altered in two-field context. 
In this work we propose a two-field freeze-in dark matter model through Higgs portal.
The observed dark matter relic abundance is obtained by a decay of scalar mediator thermalized in the early Universe.
While there is a lack of direct dark matter signals,
the scalar mediator is in the reach of HL-LHC either through vector boson fusion or Mono-Z channel.
Within allowed scalar mass window of 10-50 GeV,
we use improved cuts to derive both $2\sigma$ exclusion and $5\sigma$ discovery limits,
depending on the value of Higgs portal coupling.
If verified, this scalar mediator signal allows us to infer the freeze-in dark matter.

\end{abstract}

\renewcommand{\thefootnote}{\arabic{footnote}}
\setcounter{footnote}{0}
\thispagestyle{empty}
\vfill
\newpage
\setcounter{page}{1}

\tableofcontents

\section{Introduction}
Thermal dark matter (DM) has been severely challenged by a lack of conclusive signals in direct, collider and indirect detection experiments. 
For example, the direct detection limits have been close to the neutrino floor within a certain DM mass range; see e.g., \cite{Arcadi:2024ukq, Cirelli:2024ssz} for recent reviews.
Inspired by those experimental results, 
there is a growing interest in non-thermal DM produced in a few ways other than freeze-out mechanism such as freeze-in mechanism \cite{Hall:2009bx}.\footnote{An alternative mechanism not so precisely understood is through gravitational portal (during) after inflation, as the inflaton sector is yet conclusive.}
Single-field examples of freeze-in DM (FIDM) include axion-like scalar \cite{Langhoff:2022bij,Jain:2024dtw}, 
sterile neutrino fermion \cite{Asaka:2005cn,Becker:2018rve,Chianese:2018dsz,Datta:2021elq}  and dark photon \cite{Holdom:1985ag}, 
some of which are well motivated by new physics beyond Standard Model (SM).
A common feature of these FIDM models is  the absence at DM direct and collider detection experiments, 
as a result of feeble DM coupling to the SM sector. 
Within DM mass range larger than MeV scale,
only cosmic \cite{Xu:2024uas,Xu:2024cof} or astrophysical \cite{Cadamuro:2011fd,Vertongen:2011mu,Essig:2013goa,Cohen:2016uyg} ray data has  touched on this type of DM parameter space so far.

\begin{figure}
\centering
\includegraphics[width=12cm,height=3.2cm]{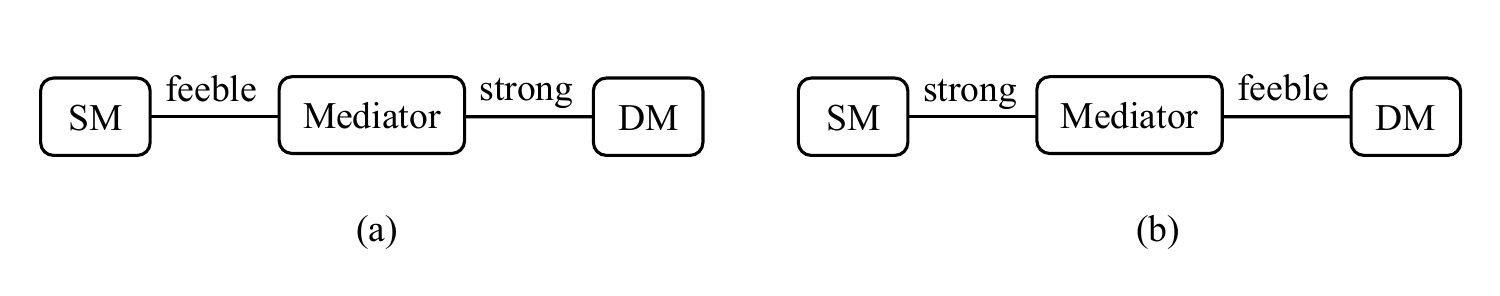}
\centering
\caption{An illustration of the two-field FIDM model classifications.}
\label{scheme}
\end{figure}

The situation changes in two-field FIDM.
A two-field FIDM, which contains DM and its force mediator, can be constructed in two different ways as shown in fig.\ref{scheme}, 
depending on how to assign the feeble interaction. Explicitly, 
\begin{itemize}
\item In the first case $(a)$ one chooses a feeble interaction between the force mediator and SM sector but a strong coupling between the DM and force mediator. 
A representative example is dark photon mediated DM,
which has been used to explain EDGES 21-cm anomaly \cite{Kovetz:2018zan,Liu:2019knx} and small-scale problem \cite{Hambye:2018dpi} respectively.
\item In the second case $(b)$ one adopts a strong interaction between the force mediator and SM sector but a feeble coupling between the DM and force mediator.
In this situation, collider-friendly FIDM models \cite{Hessler:2016kwm,Ghosh:2017vhe,Calibbi:2018fqf,Belanger:2018sti} can be constructed.
\end{itemize}

In this work, we investigate a new two-field FIDM of the second type with the following Lagrangian
\begin{eqnarray}\label{Lag}
\mathcal{L}_{D}=\bar{\psi}\left(i\gamma^{\nu}\partial_{\nu}-m_{\psi}\right)\psi+\frac{1}{2}\partial_{\mu} s\partial^{\mu}s-\frac{\mu^{2}_{s}}{2}s^{2}-\kappa s^{2}\mid H\mid^{2}-\lambda \bar{\psi}\psi s,
\end{eqnarray}
where fermion $\psi$ and real singlet scalar $s$ is the DM and force mediator respectively,
and $H$ is the SM Higgs doublet.
We impose a $Z_{2}\times Z'_{2}$ dark parity on the dark sector, under which  
\begin{eqnarray}\label{parity}
Z_{2}: \psi\rightarrow -\psi; ~~~~~ Z'_{2}:~~s \rightarrow -s,
\end{eqnarray}
and the SM sector is even.
In eq.(\ref{Lag}), we assume the breaking of $Z_{2}\times Z'_{2}\rightarrow Z_2$ by the renormalizable Yukawa interaction in the dark sector among others,\footnote{We neglect $Z'_2$-preserving $s^{4}$ term that does not affect either DM or collider phenomenological analysis.
In addition, we ignore $Z'_2$-breaking $s\mid H\mid^{2}$ term leading to tiny mixing effect between $s$ and Higgs, whose magnitude can be of the same order as $\lambda$.}
which suggests a small $\lambda$ being by following the argument of naturalness \cite{tHooft:1979rat};
and the scalar mass squared reads as $m^{2}_{s}=\mu_{s}^{2}+\kappa \upsilon^{2}\geq \kappa \upsilon^{2}$ with the electroweak scale $\upsilon=246$ GeV.
Unlike in the  Higgs-portal scalar DM \cite{Silveira:1985rk,McDonald:1993ex,Burgess:2000yq} based on freeze-out mechanism,
where only a narrow scalar mass window near a half of the Higgs mass is not excluded \cite{Cline:2013gha,Han:2015hda},
the FIDM due to a small $\lambda$ in eq.(\ref{Lag}) allows us to reopen the scalar mass window above GeV scale. 
Within this mass window, the scalar mediator which only decays to a pair of DM particles contributes to observable missing energy \cite{Han:2016gyy} at the LHC for $\kappa$ being not far away from unity, making our model differ from previous two-field models \cite{Hessler:2016kwm,Ghosh:2017vhe,Calibbi:2018fqf,Belanger:2018sti} where the mediator therein decays to either SM lepton or hadron final states.

The rest of this paper is organized as follows. 
In Sec.\ref{DMP} we discuss DM phenomenology including DM production, scattering off SM protons, and annihilation into SM particles.
Although there is a lack of DM signals in traditional DM detection experiments, 
the scalar mediator can leave footprints in the LHC, depending on the value of Higgs portal coupling $\kappa$.
Sec.\ref{LHCP} is devoted to investigate high luminosity (HL)-LHC reaches of the scalar mediator in allowed scalar mediator mass range of $10-60$ GeV through the Higgs portal induced vector boson fusion (VBF) and Mono-Z channel, where both $2\sigma$ exclusion and $5\sigma$ discovery limits are presented.
 Appendix.\ref{cuts} shows details of how to improve cuts than those of \cite{Han:2016gyy}.
Finally, we conclude in Sec.\ref{CON}.

\section{Dark matter phenomenology}
\label{DMP}
In this section we begin with the DM phenomenology in DM parameter space composed of $\{m_{\psi}, m_{s}, \lambda\}$ in eq.(\ref{Lag}) with a negligible dependence on $\kappa$ for $\kappa$ larger than  $\sim 10^{-6}$ required by the freeze-out mechanism.

\subsection{Relic abundance}
With $\kappa$ being large enough, the scalar mediator keeps thermal equilibrium with SM thermal bath. 
In contrast, for a tiny $\lambda$ the DM is produced in the early Universe through the freeze-in mechanism with Boltzmann equation 
\begin{eqnarray}\label{Boltz}
\dot{n}_{\psi}+3Hn_{\psi}=\mathcal{C},
\end{eqnarray}
where $n_{\psi}$ is the DM number density, $H$ is the Hubble rate due to cosmic expansion, and $\mathcal{C}$ is the ``collision" term mainly arising from  (a) decay $s\rightarrow \psi\bar{\psi}$ in the DM mass range of $m_{s}>2m_{\psi}$ considered here
and (b) two-body annihilation $h~s\rightarrow \psi\bar{\psi}$ via the $\kappa$ coupling.

\begin{figure}
\centering
\includegraphics[width=12cm,height=8cm]{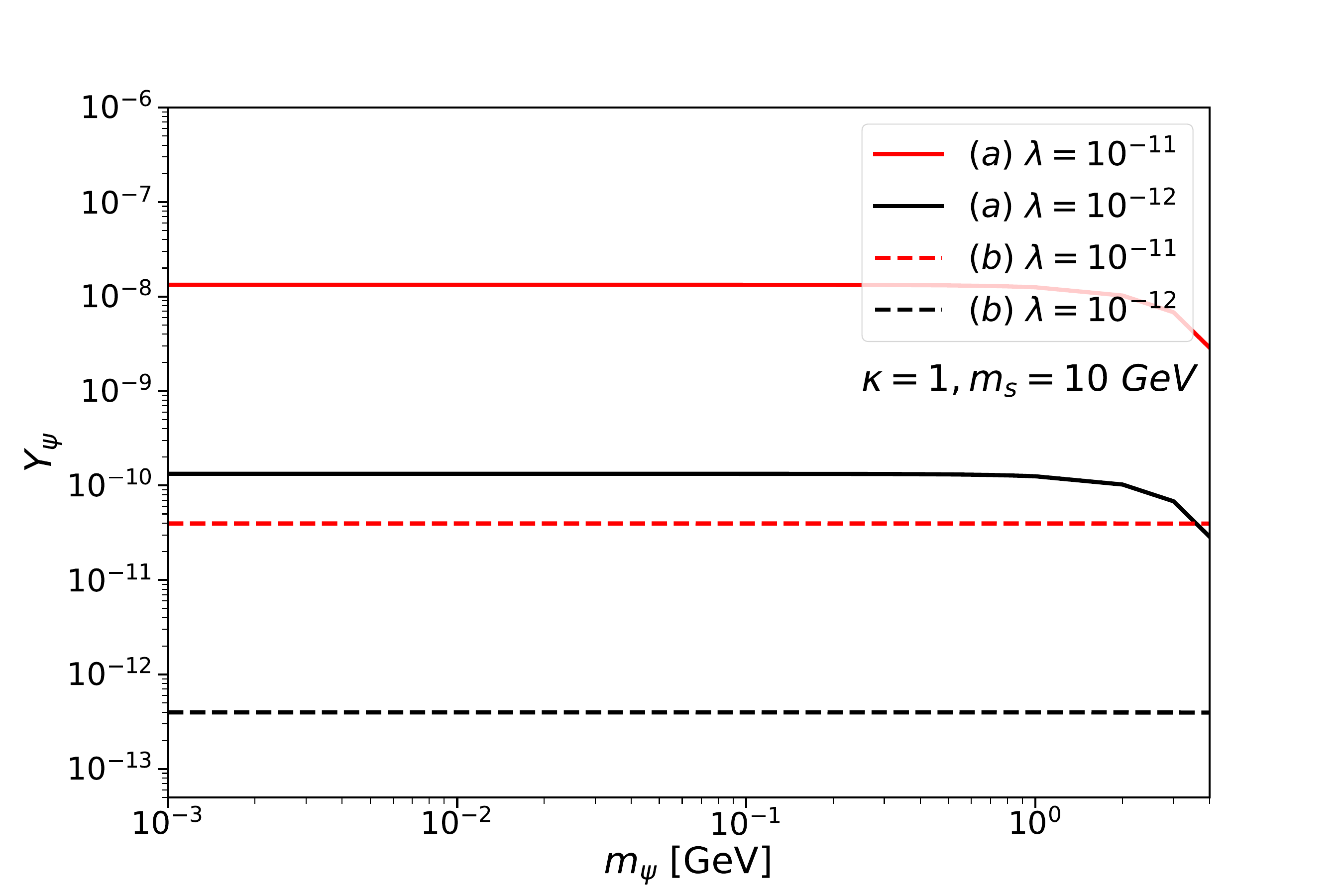}
\centering
\caption{An illustration of the individual contribution to $Y_{\psi}$ as function of $m_{\psi}$ for a set of fixed values $\kappa=1$, $\lambda=\{10^{-12}, 10^{-11}\}$ and $2m_{\psi}<m_{s}=10$ GeV,
where $(a)$ and $(b)$ is shown in solid and dashed, respectively.}
\label{yield}
\end{figure}

To see which process dominates the $\mathcal{C}$ term in eq.(\ref{Boltz}),
we present in fig.\ref{yield} the individual contribution to DM yield $Y_{\psi}\equiv n_{\psi}/S$ as function of $m_{\psi}$ with $S$ the entropy of SM thermal bath, for a set of fixed values $\kappa=1$, $\lambda=\{10^{-12}, 10^{-11}\}$ and $2m_{\psi}<m_{s}=10$ GeV.
In this plot (a) and (b) is shown in solid and dashed, respectively.
\begin{itemize}
\item The decay induced contribution to $Y_{\psi}$ is consistent with an analytic approximation \cite{Hall:2009bx}
\begin{eqnarray}\label{yyield}
Y_{\psi}\approx \frac{135}{4\pi^{3}\cdot 1.66 g_{s,*}\sqrt{g_{\rho,*}}}\frac{M_{P}\Gamma_{s}}{m^{2}_{s}},
\end{eqnarray}
where $g_{s,*}$ and $g_{\rho,*}$ are the number of degrees of freedom in entropy and energy density respectively,
$M_P$ is the non-reduced Planck mass, and the scalar mediator decay width
\begin{eqnarray}\label{swid}
\Gamma_{s}(s\rightarrow \psi\bar{\psi})=\frac{\lambda^{2}}{8\pi}m_{s}\left(1-\frac{4m^{2}_{\psi}}{m^{2}_{s}}\right)^{3/2}.
\end{eqnarray}
\item The annihilation induced contribution to $Y_{\psi}$ is proportional to $\sim \lambda^{2}\kappa^{2}$.
As seen in fig.\ref{yield}, this contribution is smaller than the decay induced one by more than two orders,
which is further reduced by choosing $\kappa$ smaller than unity. 
\end{itemize}
Therefore, in the parameter regions with $10^{-6}<\kappa\leq1$ and $2m_{\psi}<m_{s}<1$ TeV considered in this paper
the DM freeze-in production in the early Universe  is dominated by the scalar decay process.

\begin{figure}
\centering
\includegraphics[width=12cm,height=8cm]{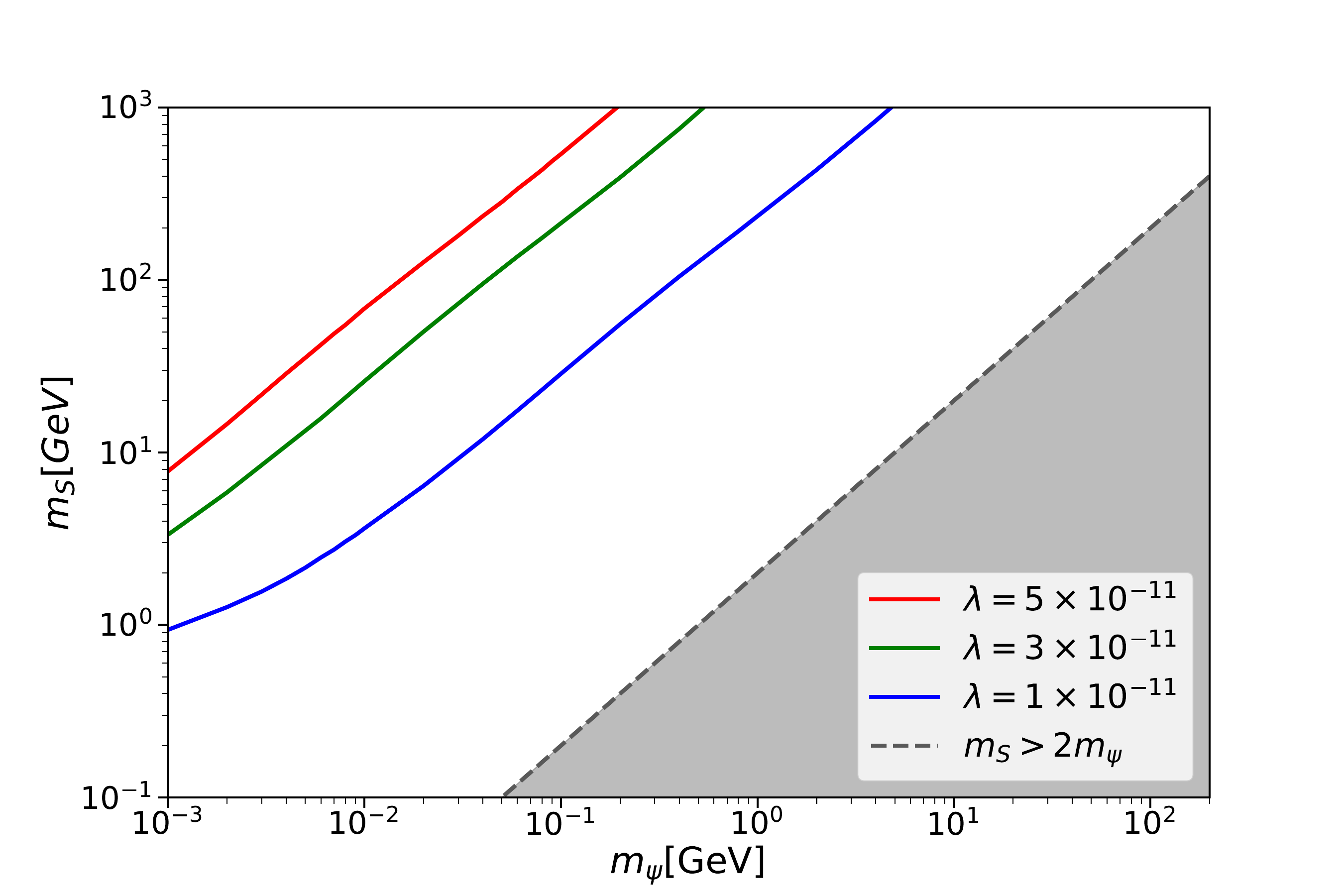}
\centering
\caption{Contours of the observed DM relic abundance projected to the plane of $m_{s}-m_{\psi}$ for various values of $\lambda$,
where the shaded region is excluded by $2m_{\psi}<m_{s}$ and the top left region with $\lambda\sim 10^{-11}$ and $m_{s}\geq 1$ GeV evades the BBN constraint on $\tau_{s}$.}
\label{dmrelic}
\end{figure}

In terms of publicly available code \texttt{micrOMEGAs6.0} \cite{Alguero:2023zol}, 
we show in fig.\ref{dmrelic} contours of the observed relic density $\Omega_{\psi}h^{2}=0.12\pm 0.001$ \cite{Planck:2018vyg}
projected to the plane of $m_{s}-m_{\psi}$ for various values of $10^{11}\lambda=\{1, 3, 5\}$,
where the shaded region is excluded by $2m_{\psi}<m_{s}$.
The numerical results of fig.\ref{dmrelic} are consistent with the analytic approximation 
\begin{eqnarray}\label{relic}
\Omega_{\psi}h^{2}=\frac{m_{\psi}Y_{\psi}S_{0}}{\rho_{\rm{crit},0}}\approx 0.1 \left(\frac{\lambda}{5\times 10^{-13}}\right)^{2}\left(\frac{m_{\psi}}{m_{s}}\right),
\end{eqnarray}
derived from eq.(\ref{yyield}), where $S_0$ is the present value of $S$ and $\rho_{\rm{crit},0}$ is the critical energy density.
Take an explicit value of $\lambda=5\times 10^{-11}$ for example,
in which case $m_{s}/m_{\psi}\sim 10^{4}$ as shown by the red curve is verified by eq.(\ref{relic}).
On the other hand, eq.(\ref{relic}) suggests that a curve with respect to a larger $\lambda$ should have a larger slope, 
which is also justified by comparing the curves in this plot.

The contours of $\Omega_{\psi}h^{2}$ in fig.\ref{dmrelic} barely depend on $\kappa$; see also eq.(\ref{relic}). 
This is true as long as $\kappa$ is larger than the threshold value $\sim 10^{-6}$ required to keep the scalar mediator in thermal equilibrium with the SM thermal bath. 
In this situation, $\kappa$ only affects the DM relic density through altering the values of $g_{s,*}$ and $g_{\rho,*}$ in eq.(\ref{yyield}) by $\sim 1\%$, which can be safely neglected.

Before closing the analysis on the DM relic density, we remind that the scalar mediator lifetime is 
\begin{eqnarray}\label{lifetime}
\tau_{s}\sim 10~\rm{sec} \left(\frac{\lambda}{10^{-12}}\right)^{-2}\left(\frac{m_{s}}{1~\rm{GeV}}\right)^{-1},
\end{eqnarray}
using eq.(\ref{swid}).
Eq.(\ref{lifetime}) shows that the lifetime is always shorter than $\sim 10^{4}$ sec, 
when the big-bang nucleosynthesis (BBN) takes place, 
in the top left region of fig.\ref{dmrelic} with $\lambda \sim 10^{-11}$ and $m_{s}\geq 1$ GeV,
which evades the BBN constraint \cite{Feng:2003uy,Hooper:2011aj}.

\subsection{Direct and indirect detection}
\begin{figure}
\centering
\includegraphics[width=7cm,height=4cm]{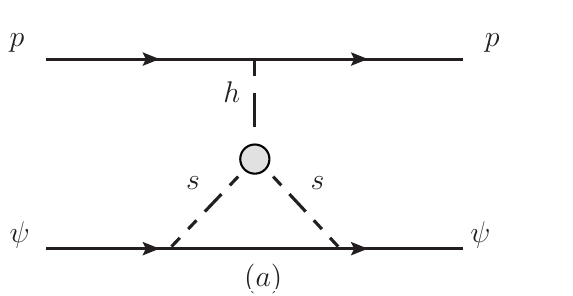}~~~~
\includegraphics[width=7cm,height=4cm]{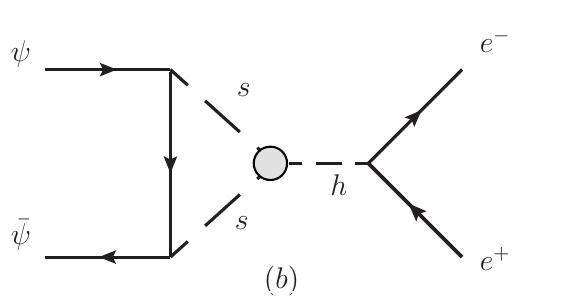}
\centering
\caption{Feynman diagrams contributing to $(a)$ spin-independent DM direct detection and (b) DM indirect detection, respectively.}
\label{DMd}
\end{figure}

In our model the DM scatters off protons or electrons by one-loop Feynman diagram.
Consider that the electron Yukawa coupling is small, 
we focus on spin-independent DM-proton cross section $\sigma_{\rm{SI}}$ as shown in diagram (a) of fig.\ref{DMd}.
It can be estimated by a non-relativistic effective operator analysis in low energy region.
Integrating out the scalar mediator in eq.(\ref{Lag}), one finds
\begin{eqnarray}\label{effL}
\mathcal{L}_{\rm{eff}}\sim \kappa\lambda^{2}\frac{m^{3}_{\psi}}{m^{4}_{s}}\psi\bar{\psi} \mid H\mid^{2}+\cdots,
\end{eqnarray}
which gives
\begin{eqnarray}\label{sigmasi}
\sigma_{\rm{SI}}\sim 10^{-86}\rm{cm}^{2}\cdot \kappa^{2}\cdot\left(\frac{\lambda}{10^{-11}}\right)^{4},
\end{eqnarray}
for the reference masses $m_{s}\sim m_{\psi}\sim 100$ GeV.
Normalized to $\kappa=1$, the value of $\sigma_{\rm{SI}}$ suppressed by $\lambda^{4}$
is roughly $\sim 42$ orders of magnitude smaller than the latest XENON1T \cite{XENON:2018voc}, 
PandaX-4T \cite{PandaX-4T:2021bab} and LZ \cite{LZ:2022lsv} bounds. 
Note,  the effective operator analysis in eq.(\ref{effL}) is valid for a rough estimate on the magnitudes of $\sigma_{\rm{SI}}$.
Similar conclusion stands for sub-GeV DM compared to the SENSEI \cite{SENSEI:2023zdf} and DAMIC\cite{DAMIC-M:2025luv} bounds on DM-electron scattering cross section.

As mentioned in the Introduction, 
so far the measurements of astrophysical or cosmic rays place the most stringent constraints on DM annihilation cross sections within the DM mass range of $m_{\psi}>1$ MeV. 
Take the $e\bar{e}$ final states as shown in diagram (b) of fig.\ref{DMd} for example. 
The DM annihilation cross section reads as 
\begin{eqnarray}\label{anncs}
\sigma_{\rm{ann}}(\psi\bar{\psi}\rightarrow e\bar{e})v_{\rm{rel}}\sim 10^{-68}\rm{cm}^{3}\rm{s}^{-1}\cdot \kappa^{2}\cdot\left(\frac{\lambda}{10^{-11}}\right)^{4},
\end{eqnarray}
for the reference mass $m_{\psi}\sim 100$ GeV by using eq.(\ref{effL}), 
where $v_{\rm{rel}}$ is the relative velocity of two incoming DM particles.  
Due to the $\lambda^{4}$ suppression, 
the value of $\sigma_{\rm{ann}}(\psi\bar{\psi}\rightarrow e\bar{e})v_{\rm{rel}}$ in eq.(\ref{anncs}) is roughly $\sim 38$ orders of magnitude smaller than current cosmic ray bounds, see e.g \cite{Xu:2024uas}.

\section{LHC phenomenology}
\label{LHCP}
While it is unlikely to probe the freeze-in DM at the DM detection facilities, 
the scalar mediator $s$ which only decays to a pair of DM particles can be probed by the LHC, 
given the $\kappa$ coupling to Higgs being large enough.
In this section, we firstly discuss current LHC limit on Higgs invisible decay, 
then turn to HL-LHC reaches of $s$ with center-of-mass energy fixed to be 14 TeV.

\subsection{Indirect detection}
For $2m_{s}<m_{h}$ with $m_h$ the observed Higgs mass, the scalar mediator contributes to the invisible decay of Higgs with decay width 
\begin{eqnarray}\label{hinv}
\Gamma^{\rm{inv}}_{h}(h\rightarrow ss)=\frac{\kappa^{2} \upsilon^{2}}{8\pi m_{h}}\sqrt{1-\frac{4m^{2}_{s}}{m^{2}_{h}}},
\end{eqnarray}
which is upper bounded as $\Gamma^{\rm{inv}}_{h}\leq 0.26~\Gamma_{h}$ at 95$\%$ CL \cite{ATLAS:2019cid} with 
the Higgs decay width $\Gamma_{h}\approx 4.15$ MeV.

\begin{figure}
\centering
\includegraphics[width=6cm,height=5cm]{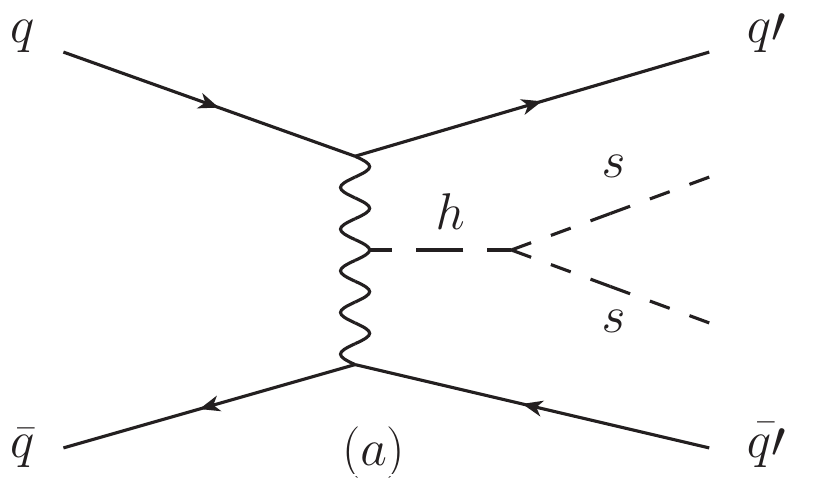}~~~~~~~~~~~
\includegraphics[width=6cm,height=5cm]{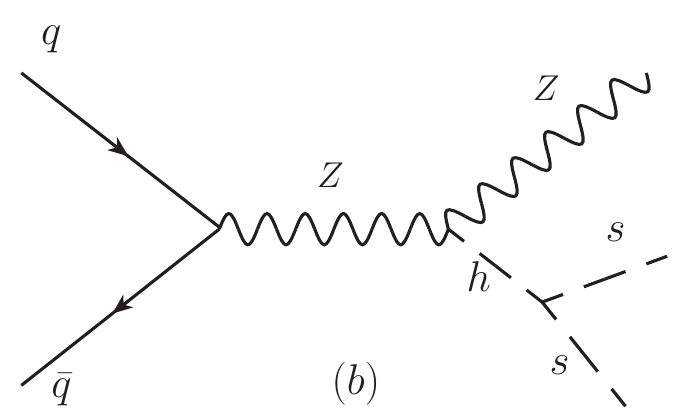}
\centering
\caption{The leading-order Feynman diagram with respect to (a) VBF and (b) Mono-Z channel at the LHC, respectively.}
\label{Feyn}
\end{figure}

\subsection{Direct detection}
Given the $\kappa$ coupling between $s$ and $H$, 
the scalar mediator can be produced through both VBF and Mono-Z channel similar to the Higgs-portal scalar DM \cite{Han:2016gyy}. 
As mentioned in the Introduction, 
unlike in \cite{Han:2016gyy} where only $m_{s}\sim m_{h}/2$ is not excluded,
our model reopens a larger mass window of $m_{s}\geq 1$ GeV.

\subsubsection{Productions}
We use FeynRules \cite{Alloul:2013bka} to generate model files prepared for MadGraph5 \cite{Alwall:2014hca}
which includes Pythia 6 \cite{Sjostrand:2006za} for parton showering and hadronazition,
and the package Delphes 3 \cite{deFavereau:2013fsa} for fast detector simulation.\footnote{Apart from such sophisticated method, new machine learning methods such as XGBoost \cite{Chen:2016} are also viable.}

\begin{figure}
\centering
\includegraphics[width=8cm,height=8cm]{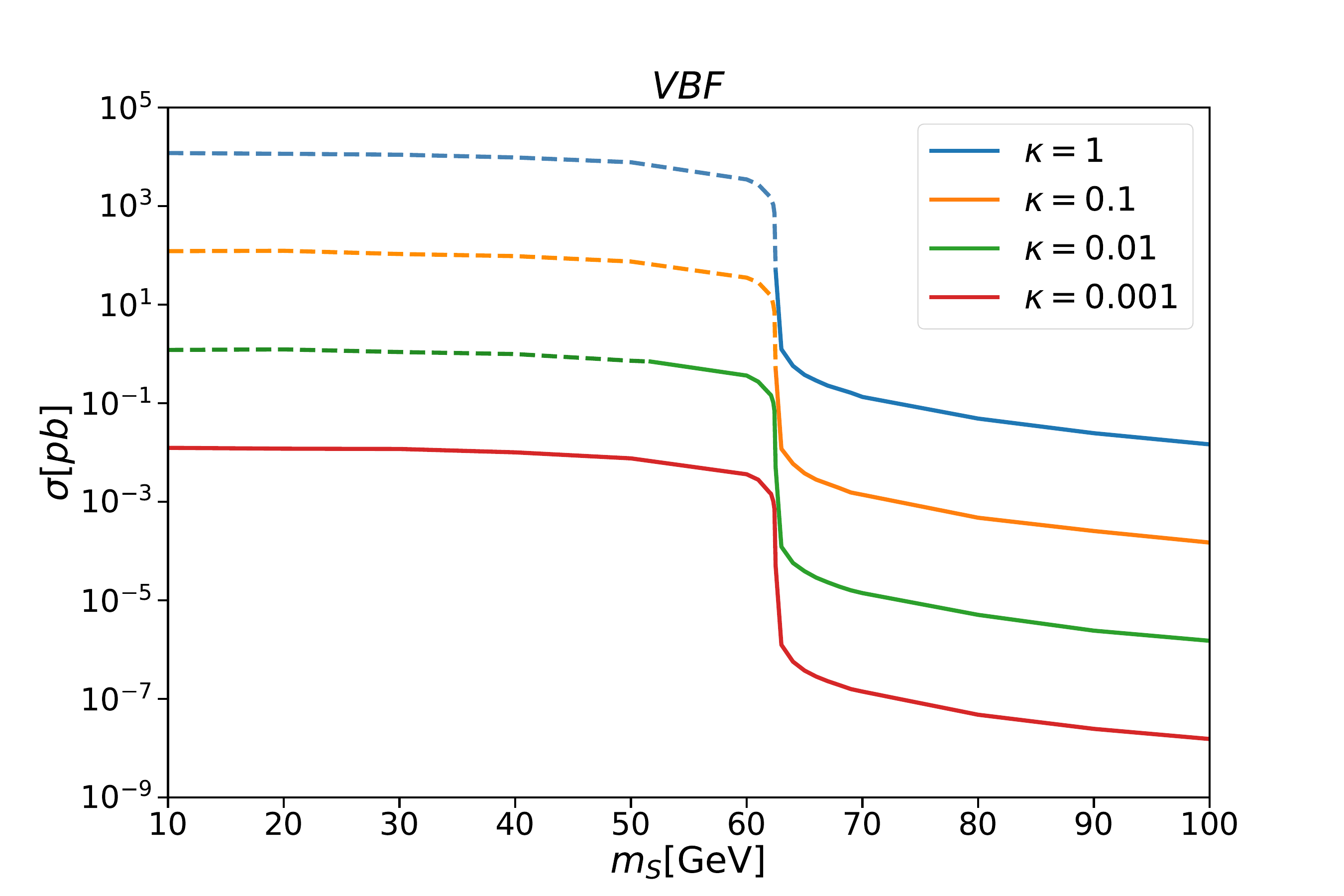}~~~~~~~
\includegraphics[width=8cm,height=8cm]{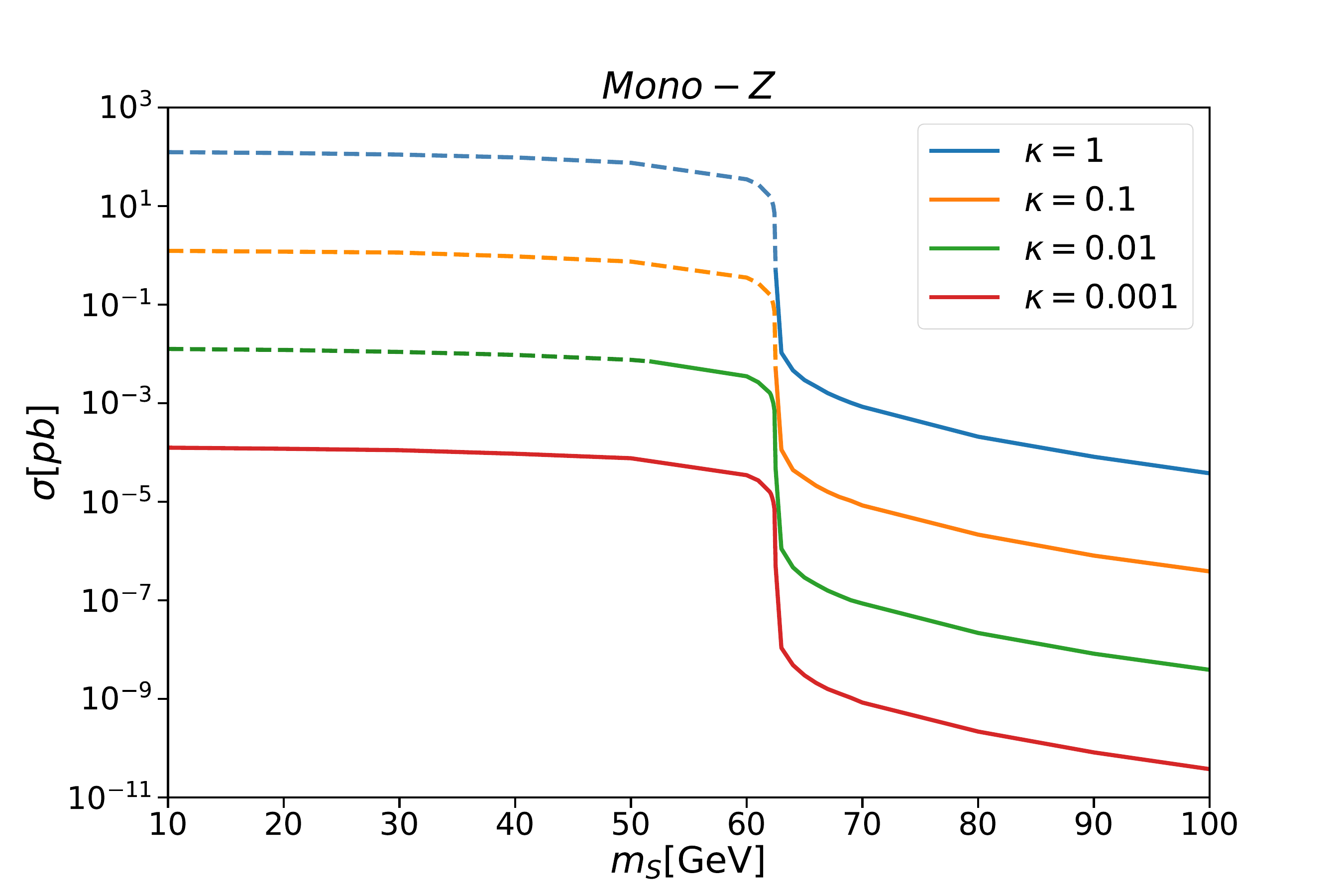}
\centering
\caption{Values of signal cross section $\sigma_{s,\rm{VBF}}$ (left) and $\sigma_{s, \rm{MZ}}$ (right) as function of $m_s$ for various values of $\kappa=\{10^{-3}, 10^{-2}, 10^{-1},1\}$ at the 14-TeV LHC, 
which have a significant decline at $m_{s}\sim m_{h}/2$. With respect to each $\kappa$, the dotted line is excluded by the current bound on $\Gamma^{\rm{inv}}_{h}$ in eq.(\ref{hinv}).}
\label{cs}
\end{figure}

$\underline{\mathbf{VBF}}$. 
As induced by mixing effect \cite{Mou:2015mdm}, 
the VBF channel corresponds to $q\bar{q}\rightarrow h^{*}+jj\rightarrow ss+jj$ as shown in the plot (a) of fig.\ref{Feyn}, 
where $q$ and $q'$ refers to SM quarks, and $s$ contributes to a large missing transverse momentum (MET).
We present in the left plot of fig.\ref{cs} the values of signal cross section $\sigma_{\rm{s,VBF}}$ as function of $m_s$ at the 14-TeV LHC, 
for various values of $\kappa=\{10^{-3}, 10^{-2}, 10^{-1},1\}$.
In this plot,  for a fixed value of $\kappa$ the magnitudes of $\sigma_{\rm{s,VBF}}$ vary over six orders from $m_{s}=10$ GeV to 100 GeV, 
which significantly decrease at $m_{s}\sim m_{h}/2$.
With respect to this VBF channel, 
the SM background arises from $W(\ell \nu)+$jets, $Z(\nu\bar{\nu})+$jets, with subdominant $t\bar{t}+$jets and QCD multi-jets neglected.
For the final states with 2 jets, the cross section $\sigma_{\rm{b,VBF}}\approx 6.4~(1.2)\times 10^{4}$ pb,
implying that a large portion of the parameter regions with $m_{s}\sim 10-100$ GeV and $\kappa\sim 10^{-3}-1$ is in the reach of HL-LHC.

$\underline{\mathbf{ Mono-Z}}$. 
The Mono-Z channel refers to $q\bar{q}\rightarrow h^{*}+Z\rightarrow ss+Z(\ell\bar{\ell})$ as shown in the plot (b) of fig.\ref{Feyn}.
The right plot of fig.\ref{cs} shows the values of signal cross section $\sigma_{\rm{s,MZ}}$ as function of $m_s$ at the 14-TeV LHC for various values of $\kappa=\{10^{-3}, 10^{-2}, 10^{-1}, 1\}$.
In this plot,
the magnitudes of $\sigma_{\rm{s,MZ}}$ have a dramatical decline at $m_{s}\sim m_{h}/2$ similar to the left plot of fig.\ref{cs}, despite being smaller. 
Regarding this Mono-Z channel, the SM background arises from $Z(\nu\bar{\nu})Z(\ell\bar{\ell})$ with an explicit value of cross section $\sigma_{\rm{b,MZ}}\approx 30$ pb,
where $\ell=e,\mu$.
Compared with the VBF channel, the mono-Z channel is sub-leading but relatively cleaner.
Likewise, a large portion of the parameter regions with $m_{s}\sim 10-100$ GeV and $\kappa\sim 10^{-3}-1$ can be also reached by the HL-LHC.

Compared to the VBF process, while gluon fusion process produces a $\sim 5$ times larger cross section of monojet signal in the mass range of $m_{s}\sim 10-50$ GeV, the efficiency of this signal selection is however much smaller \cite{ATLAS:2021kxv}, making the ability of exclusion weaker, see e.g \cite{ATLAS:2023tkt}. 
As a result, we do not consider the gluon fusion process in the following analysis.

\subsubsection{Cuts}
$\underline{\mathbf{VBF}}$. Ref.\cite{Han:2016gyy} has applied the earlier CMS cuts in \cite{CMS:2014gab} to a numerical simulation of the scalar DM via the VBF channel. 
Those cuts give a ratio of $\epsilon_{s}/\sqrt{\epsilon_{b}}\sim 5$ for a million events,
where $\epsilon_{s}$ and $\epsilon_{b}$ refers to signal and background efficiency respectively. 
In this work we make use of recently reported CMS \cite{CMS:2022qva} and ATLAS \cite{ATLAS:2022yvh} cuts to update the simulation of VBF channel. Explicitly, 
\begin{itemize}
\item First, we impose the following trigger and selection cuts:
\begin{eqnarray}\label{cut1}
\Delta R_{jj}>0,~~~~\eta_{j}<5,~~~~P_{T}^{j}>25~\rm{GeV},
\end{eqnarray}
where $\Delta R_{jj}=\sqrt{(\Delta\phi)^{2}+(\Delta\eta)^{2}}$ with $\Delta\eta$ the pseudorapidity difference between the two leading jets, 
$P_{T}^{j}$ and $\eta_{j}$ is the transverse momentum and pseudorapidity of jet $j=1,2$ respectively.
 
\item Second, we use \texttt{MadAnalysis} \cite{Conte:2012fm} to analyze the distribution of both signal and background events as function of relevant parameters in fig.\ref{vbfd} of appendix.\ref{cuts},
where $P_{T}^{j_{1(2)}}$ and $\eta_{j_{1(2)}}$ is the transverse momentum and pseudorapidity of the first (second) leading jet respectively,
$M_{jj}$ invariant mass of the two leading jets, $E_{T}^{\rm{miss}}$ the missing transverse energy.

\item Third, we choose the explicit values of those cut parameters one by one by referring the cuts used in \cite{CMS:2022qva, ATLAS:2022yvh}. 
During this process, when a new added cut such as azimuthal angle $\Delta\phi$ between the two jets does not offer an effective increase in the ratio of $\epsilon_{s}/\sqrt{\epsilon_{b}}$ but obviously reduces the number of signal events, 
we neglect that cut parameter for simplicity. 
\end{itemize}
Table.\ref{fcuts} shows the final cuts applied to the VBF channel,
in terms of which the value of $\epsilon_{s}/\sqrt{\epsilon_{b}}$ is enhanced by $\sim 3-4$ times compared to that of ref.\cite{Han:2016gyy}.

\begin{table}
\begin{center}
\begin{tabular}{cc}
\hline\hline
VBF  ~~~~~~~~~& Mono-Z \\ \hline
$N(\rm{jet})\geq 2$ &  $N(\ell)$=2,~$\ell=e,\mu$ \\
$N(\ell)=0$ &  $P_{T}^{\ell\ell}>100$ \rm{GeV} \\
$P_{T}^{j_{1}}>60$ \rm{GeV} &  $\mid M_{\ell\ell}-m_{Z}\mid>10$ \rm{GeV} \\
$\mid\eta_{j_{2}}\mid>3$  & $E_{T}^{\rm{miss}}>180$ \rm{GeV}   \\
$E_{T}^{\rm{miss}}>100$ \rm{GeV}  & $\mid E_{T}^{\rm{miss}}-P_{T}^{\ell\ell}\mid/P_{T}^{\ell\ell}<1$ \\ 
$M_{jj}>1100$ \rm{GeV} &~~~~ $\Delta\phi(\overrightarrow{P}^{\ell\ell}_{T}, \overrightarrow{P}_{T}^{\rm{miss}})>2.5$ rad \\
$\Delta R_{jj}>5.5$ & $\Delta R_{\ell \ell}<1.2$  \\
\hline \hline
\end{tabular}
\caption{Cuts adopted for numerical simulations of the VBF (left) and Mono-Z (right) channel, respectively.}
\label{fcuts}
\end{center}
\end{table}

$\underline{\mathbf{Mono-Z}}$. 
Using the earlier CMS cuts in \cite{CMS:2015rjz} that give a ratio of $\epsilon_{s}/\sqrt{\epsilon_{b}}\sim 15$ for a million events,
ref.\cite{Han:2016gyy} also analyzed the scalar DM reach via the Mono-Z channel at the HL-LHC.
Now we repeat this analysis for the scalar mediator by referring to updated cuts in \cite{CMS:2017ret, CMS:2017nxf, CMS:2020ulv}. 
To do so,
\begin{itemize}
\item First,  as in eq.(\ref{cut1}) we impose the following trigger and selection cuts:
\begin{eqnarray}\label{cut2}
\Delta R_{\ell \ell}>0.4,~~~~\eta_{\ell}<2.5, ~~~~P_{T}^{\ell}> 10~\rm{GeV},
\end{eqnarray}
where $\Delta R_{\ell \ell}$ is the separation between two leptons, 
$P_{T}^{\ell}$ and $\eta_{\ell}$ is the transverse momentum and pseudorapidity of lepton $\ell=1,2$ respectively.
 
\item Second, we pick cut parameters specified for the lepton pair $\ell\bar{\ell}$ with $\ell=e,\mu$ and neglect those related to $\ell=\tau$ and jets.
We use \texttt{MadAnalysis} \cite{Conte:2012fm} to carry out the distribution of both signal and background events as function of the selected cuts in fig.\ref{mzd} of appendix.\ref{cuts},
where $P_{T}^{\ell\ell}$ is the dilepton transverse momentum, $M_{\ell\ell}$  the dilepton mass,
$\Delta\phi(\overrightarrow{P}^{\ell\ell}_{T}, \overrightarrow{P}_{T}^{\rm{miss}})$ the azimuthal separation between dilepton and missing momentum, 
and $\Delta R_{\ell \ell}$ the separation between two leptons. 

\item Third, repeating the same process as in the VBF channel, we choose the explicit values of those cut parameters one by one. 
\end{itemize}
Table.\ref{fcuts} shows the final cuts applied to the Mono-Z channel, in term of which one obtains an enhancement on the ratio of $\epsilon_{s}/\sqrt{\epsilon_{b}}$ by $\sim 2-3$ times compared to ref.\cite{Han:2016gyy}.

\subsubsection{Sensitivities of HL-LHC}
Collecting the previous results about the cross sections and efficiencies, 
we now present the $2\sigma$ exclusion ($5\sigma$ discovery) defined as 
\begin{eqnarray}\label{signif}
\frac{S}{\sqrt{B}}\left(\frac{S}{\sqrt{S+B}}\right)=2~(5),
\end{eqnarray}
where $S$ and $B$ is the number of signal and background events, respectively.
In this work we neglect systematic uncertainties of order a few percent level such as those in parton showers and the proton PDFs, see \cite{ATLAS:2022yvh,CMS:2020ulv} for details.

\begin{figure}
\centering
\includegraphics[width=15cm,height=12cm]{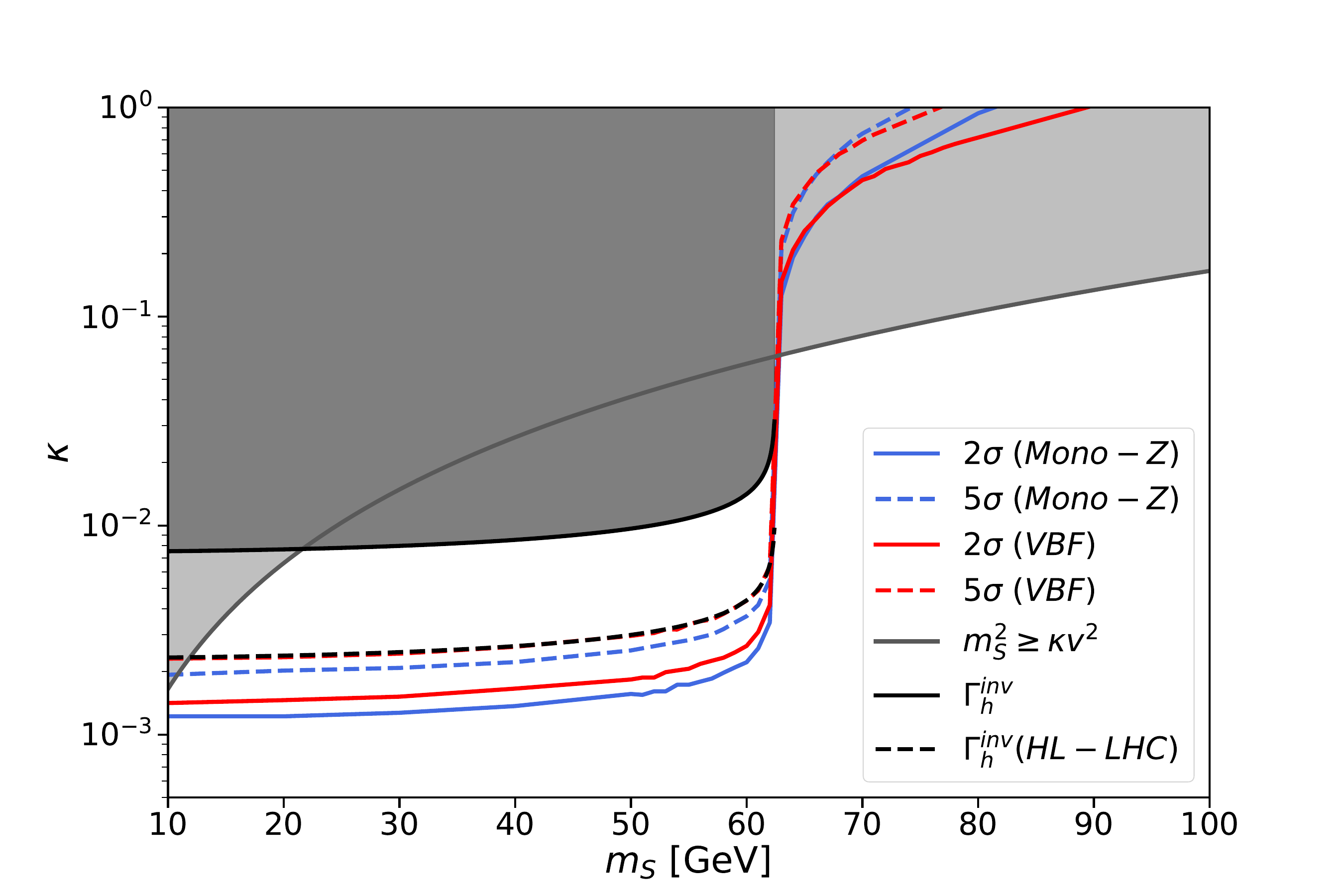}
\centering
\caption{Sensitivities of HL-LHC with $2\sigma$ exclusion and $5\sigma$ discovery with respect to the integrated luminosity $\mathcal{L}=3$ ab$^{-1}$ for the VBF (red) and Mono-Z (blue) channel respectively, 
which are projected to the plane of $m_{s}-\kappa$ evading the BBN constraint on $\tau_{s}$ in eq.(\ref{lifetime}).
Here, the shaded regions are excluded by the current LHC bound on $\Gamma^{\rm{inv}}_{h}$ (in black) and the scalar mass bound (in gray),
and the future HL-LHC limit $\Gamma^{\rm{inv}}_{h}\leq 2.5\%\Gamma_{h}$ \cite{Cepeda:2019klc} (in black dashed) is shown for comparison.}
\label{final}
\end{figure}

$\underline{\mathbf{VBF}}$. The red curves of fig.\ref{final} shows the $2\sigma$ exclusion and $5\sigma$ discovery of HL-LHC projected to the plane of $(m_{s}, \kappa)$ for the integrated luminosity $\mathcal{L}=3$ ab$^{-1}$,
where significant changes occur at $m_{s}\sim m_{h}/2$ as seen in fig.\ref{cs}. 
In this plot, the black and gray shaded region are excluded by the bound on $\Gamma_{h}^{\rm{inv}}$ in eq.(\ref{hinv}) and the scalar mediator mass bound $m^{2}_{s}\geq \kappa \upsilon^{2}$ respectively,
pointing to an allowed parameter region with $1~\rm{GeV}<m_{s}<m_{h}/2$ and $\kappa\leq 7.5\times 10^{-3}$.
In this region, the range of $\kappa\geq 2.4~(1.5)\times 10^{-3}$ with respect to $m_{s}\sim 10-60$ GeV
can be discovered (excluded), 
implying a stronger ability of exclusion than the future HL-LHC limit on the Higgs invisible decay width $\Gamma^{\rm{inv}}_{h}\leq 2.5\%\Gamma_{h}$ \cite{Cepeda:2019klc}.

$\underline{\mathbf{Mono-Z}}$. The blue curves of fig.\ref{final} presents the $2\sigma$ exclusion and $5\sigma$ discovery of HL-LHC projected to the plane of $(m_{s}, \kappa)$ for $\mathcal{L}=3$ ab$^{-1}$,
where both the significant changes in these curves and the shaded regions are the same as above.
Compared to the VBF process and the future HL-LHC limit on $\Gamma^{\rm{inv}}_{h}$, the Mono-Z process provides the strongest signal reaches.
Explicitly, the range of $\kappa\geq 2.0~(1.25)\times 10^{-3}$ with respect to $m_{s}\sim 10-60$ GeV
can be discovered (excluded).

Once detected by the HL-LHC either through the VBF or Mono-Z channel, 
the scalar signal within the mass window\footnote{This scalar mediator mass window does not overlap with the narrow mass window of thermal scalar DM.} of $m_{s}\sim 10-50$ GeV  favors an existence of the non-thermal DM as studied in Sec.\ref{DMP}, whose mass can be inferred in terms of fig.\ref{dmrelic} with respect to an explicit value of $\kappa$.

\section{Conclusion}
\label{CON}
In this work we have revisited the Higgs-portal DM scenario with the DM being non-thermal instead of thermal as previously considered in the literature. 
Specifically, we have realized this idea in the two-field context, 
where one is the DM and the other is the scalar mediator coupled to Higgs,
with the feeble interaction between them as a result of the $Z'_2$-parity breaking.

We have divided the model phenomenology analysis into two parts.
From the perspective of DM phenomenology, the DM obtains the observed relic density mainly through the decay of scalar mediator thermalized with the SM thermal bath, 
and has negligible spin-independent scattering and annihilation cross sections as expected in a typical FIDM model.
From the perspective of LHC phenomenology, 
for the Higgs portal coupling being not too far away from unity 
the scalar mediator can be detected by the LHC.
Consider that $m_{s}$ larger than $\sim$ GeV scale is favored by the BBN constraint,
we have reported the HL-LHC reaches of the scalar mediator within the allowed mass window of $m_{s}\sim 10-50$ GeV through both the VBF and Mono-Z channel. 
If verified,  this scalar signal allows us to infer the existence of a non-thermal DM such as $\psi$ proposed here,
as interpreting it as the thermal DM has been excluded.

Our study, together with the previous works \cite{Hessler:2016kwm,Ghosh:2017vhe,Calibbi:2018fqf,Belanger:2018sti}, 
serves as an illustration of rich collider phenomenology in two-field FIDM, compared to the single-field situation.
Instead of the Higgs portal considered here, 
one can use neutrino or vector portal to construct alternative two-field FIDM models,
following our classifications in the Introduction.
Different portal may provide unique signals.
We left this point for future work.

\appendix
\section{Selection of cuts}
\label{cuts}

\begin{figure}
\centering
\includegraphics[width=15cm,height=21cm]{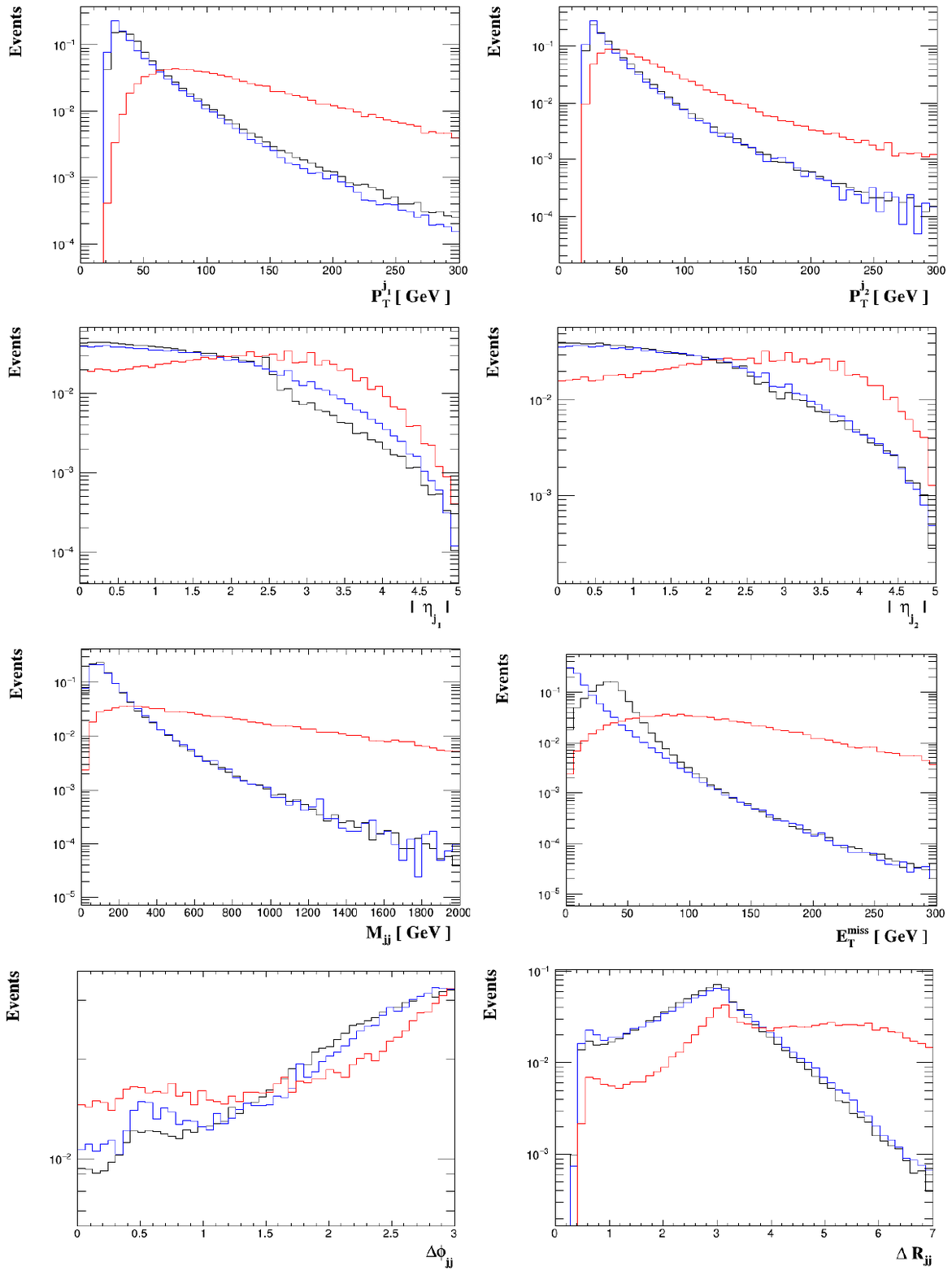}
\centering
\caption{Distribution of the VBF signal and background events,
where the signal, W+\rm{jets} and Z+\rm{jets} is shown in red, black and blue respectively.}
\label{vbfd}
\end{figure}

After imposing the cuts in eq.(\ref{cut1}), fig.\ref{vbfd} shows the distribution of VBF signal and background events as function of $p_{T}^{j_{1(2)}}$,  $\mid\eta_{ j_{1(2)}}\mid$, $E^{\rm{miss}}_{T}$, $M_{jj}$, $\Delta \phi_{jj}$ and $\Delta R_{jj}$ respectively,
which illustrates the cut values chosen for the VBF channel in Table.\ref{fcuts}.
For example, imposing $\Delta \phi_{jj}$ or $P_{T}^{j_{2}}$ indeed increases the value of $\epsilon_{s}/\sqrt{\epsilon_{b}}$ more or less but also leads to a serious suppression on the number of signal events simultaneously.
On the contrary, other cuts such $P_{T}^{j_{1}}>60$ GeV can do so with the suppression being affordable. 

\begin{figure}
\centering
\includegraphics[width=15cm,height=17cm]{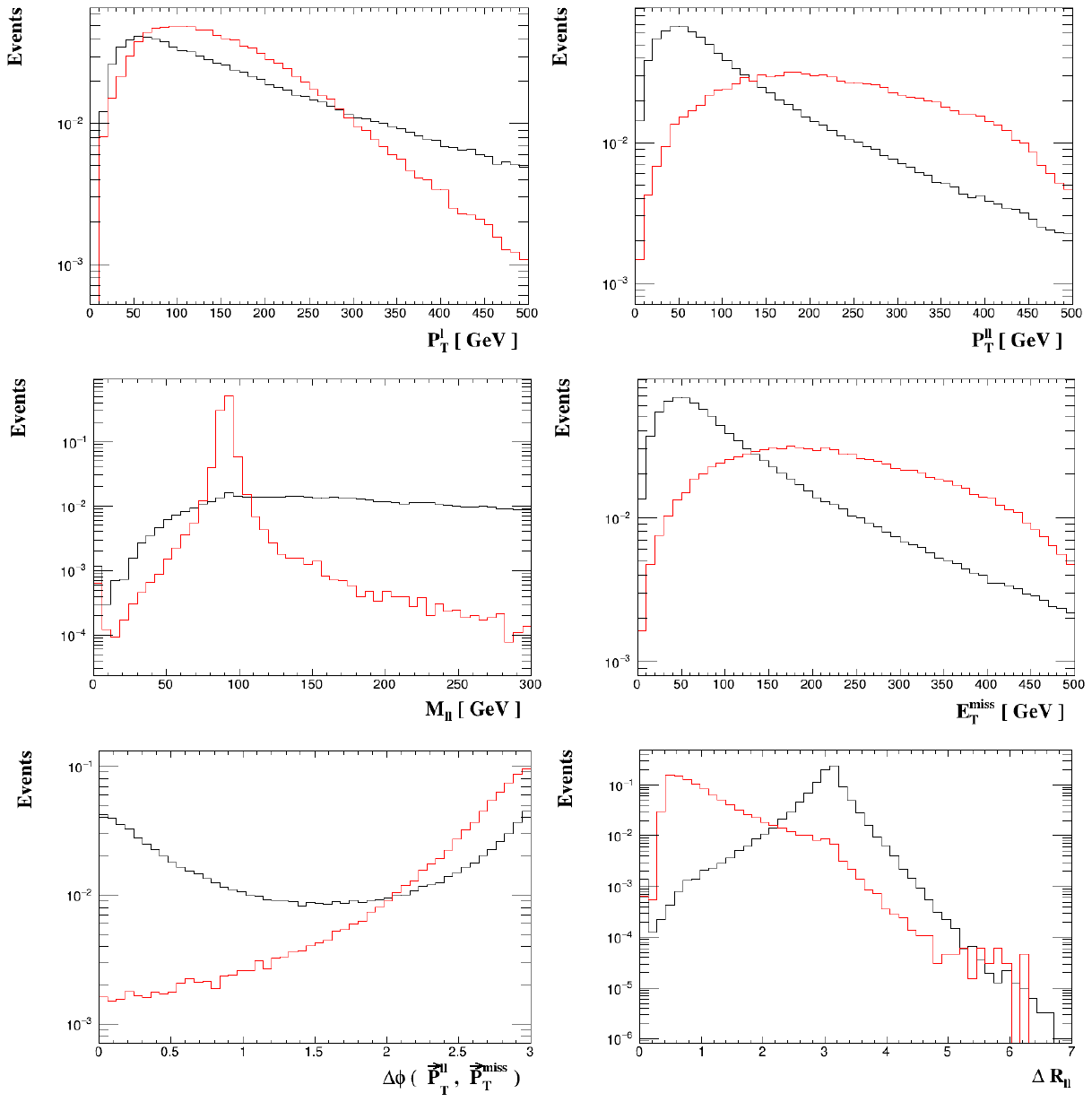}
\centering
\caption{Distribution of the  Mono-Z signal and background events,
which is shown in red and black respectively.}
\label{mzd}
\end{figure}

Likewise, after imposing the cuts in eq.(\ref{cut2}), 
fig.\ref{mzd} shows the distribution of Mono-Z signal and background events as function of $P_{T}^{\ell}$, $P^{\ell\ell}_{T}$, $M_{\ell\ell}$, $E^{\rm{miss}}_{T}$, $\Delta\phi(\overrightarrow{P}^{\ell\ell}_{T}, \overrightarrow{P}_{T}^{\rm{miss}})$ and $\Delta R_{\ell \ell}$ respectively, 
which explains the cut values chosen for the Mono-Z channel in Table.\ref{fcuts}.
For example, imposing $P_{T}^{\ell}$ is unable to increase the value of $\epsilon_{s}/\sqrt{\epsilon_{b}}$ in an efficient way as explained above, 
but other cuts such as $P_{T}^{\ell\ell}>150$ GeV can do so.

\end{document}